\documentclass[article, twocolumn,
   aps, pra,
  amsmath,amssymb,
  longbibliography,
  ]{revtex4-2}
\usepackage{graphicx,color}
\usepackage{amsmath}
\usepackage{natbib}
\usepackage{epsfig}
\begin{document}

\title{Vibrational model of entropy in dense two-dimensional fluids}

\author{S. A. Khrapak}
\email{sergey.khrapak@gmx.de}

\affiliation{Joint Institute for High Temperatures, Russian Academy of Sciences, 125412 Moscow, Russia}

\begin{abstract}
A vibrational paradigm of atomic dynamic in dense fluids is known to provide useful insight on the transport and thermodynamic properties of fluids in three dimensions. In this paper, a vibrational model is generalized to describe the excess entropy of two-dimensional (2D) fluids. A simple practical implementation of this model is demonstrated to deliver accurate results for various systems, such as one-component plasmas with Coulomb and logarithmic interactions, a 2D fluid of dipole particles, and a 2D Yukawa fluid. The applicability limits, relevance to three-dimensional fluids, relations to other 2D phenomena, and potential practical applications are briefly discussed.     
\end{abstract}

\date{\today}

\maketitle

\section{Introduction}

Recently, there has been renewed interest in developing models for the thermodynamics and transport properties of the liquid state~\cite{TrachenkoBook}.  
One of the related developments is the phonon theory of liquid thermodynamics~\cite{TrachenkoPRB2008,
BolmatovSciRep2012,BolmatovAP2015,TrachenkoRPP2015,BolmatovJPCL2022,LiuPRB2025}. The theory stems from the ideas discussed by Frenkel~\cite{FrenkelBook} that dense liquids can be approached from a solid-state perspective. Combining a similarity between the high-frequency elastic properties of liquids and solids with the Debye vibrational density of states approximation allows the calculation of the heat capacities of various liquids, which are often in good agreement with the experimental results~\cite{TrachenkoPRB2008,
BolmatovSciRep2012,BolmatovAP2015,TrachenkoRPP2015,BolmatovJPCL2022,LiuPRB2025}.  

Based largely on the same qualitative ideas that the atomic dynamics in dense liquids and fluids is dominated by solid-like vibrations about temporary local equilibrium positions, a vibrational model of entropy has been proposed~\cite{KhrapakJCP2021,KhrapakPhysRep2024,KhrapakPRE09_2024}. In this approach a solid-like expression for the entropy is complemented by the so-called communal entropy, reflecting the possibility of atoms to diffuse on long time scales. The approach yields reasonable prediction of the entropy of simple model fluids with repulsive interactions, such as a one-component plasma, inverse power-law potentials, Yukawa fluids, and even hard-sphere fluids~\cite{KhrapakJCP2021,KhrapakPhysRep2024,KhrapakPRE09_2024}.   

So far, attention has been focused on conventional fluids in three dimensions. The purpose of this paper is to generalize the vibrational paradigm of fluid dynamics to the calculation of the excess entropy of two-dimensional (2D) fluids. A simple quantitative tool is proposed to estimate the entropy of dense 2D fluids. We test the accuracy of this approach using several popular 2D fluid models such as the one-component plasma (OCP) with logarithmic and Coulomb interactions, a 2D fluid of dipoles, and a 2D Yukawa fluid. 

\section{Model}

Following Henchman~\cite{HenchmanJCP2003} we approximate the partition function of a fluid using the partition function of a classical Einstein crystal. Each particle is contained in a cell and is moving in a symmetric harmonic potential
\begin{equation}\label{harmonic}
u(r)=u_0+\frac{1}{2}m\omega_{\rm E}^2r^2,
\end{equation}
where $u$ is the energy, $u_0$ is the minimum energy in the center of the harmonic cell, $m$ is the particle mass, $\omega_{\rm E}$ is the frequency of the restoring force, known as the Einstein frequency, and $r$ is the distance from the center. The configurational part of the single cell partition function can be evaluated by integrating the Boltzmann factor of the potential:
\begin{equation}
z= \int e^{-u(r)/T}d{\bf r}=\left(\frac{2\pi T}{m\omega_{\rm E}^2}\right)e^{-u_0/T},
\end{equation}
where the temperature $T$ is expressed in energy units ($\equiv k_{\rm B}T$). The configurational integral for the entire system consisting of $N$ particles is the product $z^N$. The full partition function ${\mathcal Z}$ should also include the kinetic terms so that
\begin{equation}
{\mathcal Z}=z^N\left(\frac{mT}{2\pi\hbar}\right)^N,
\end{equation}
where $\hbar$ is the Planck constant. The Helmholtz free energy can now be evaluated from
\begin{equation}
F=-T\ln{\mathcal Z}.
\end{equation}
The Helmholtz free energy is related to the entropy via the thermodynamic identity $F=U-TS$~\cite{LandauStatPhys}. The internal energy is $U=Nu_0+2NT$, where the second term $2NT$ is equally distributed between the average kinetic energy and the vibrational energy (in 2D geometry). The entropy becomes
\begin{equation}
S=2N+2N\ln\left(\frac{T}{\hbar\omega_{\rm E}}\right)
\end{equation}
To obtain the excess entropy, we should subtract the entropy of an ideal gas at the same density and temperature. The latter is given by the 2D analogue of the Sackur-Tetrode equation~\cite{SavaraJPCC2016}
\begin{equation}
S_{\rm id}=2N+N\ln\left(\frac{mT\Delta^2}{2\pi\hbar^2}\right),    
\end{equation}
where $\Delta=n^{-1/2}$ is the characteristic distance between particles. The last step is to add the so-called ``communal entropy'', equal to $S_{\rm comm}=N$ to the excess entropy obtained. The existence of this contribution in the fluid state was suggested in
early work on cell theory~\cite{HirschfelderJCP1937}. Communal entropy was attributed to the fact that the atoms move more freely in a fluid than in a solid, can continuously interchange places and can therefore share the whole available volume. Although the concept of communal entropy has been subsequently criticized~\cite{RiceJCP1938,FrenkelBook}, recent calculations demonstrate that it represents an essential ingredient to achieve numerical precision for the entropy of simple 3D fluids~\cite{KhrapakJCP2021,KhrapakPRE09_2024}.
The final expression for the excess entropy, which will be tested in the remainder of this paper, is
\begin{equation}\label{VibrationalEntropy}
s_{\rm ex}=\frac{S_{\rm ex}}{N}=1-\ln\left(\frac{m\omega_{\rm E}^2\Delta^2}{2\pi T}\right).    
\end{equation}

\section{Validation}

The expression derived for excess entropy will now be tested on four 2D model fluids for which accurate thermodynamic data are available. Two of them are often referred to as one-component plasma (OCP). One of these OCP is characterized by the conventional 3D Coulomb interaction potential $\varphi(r)=Q^2/r$, where $Q$ is the particle charge, but the motion of charged particles is restricted to a 2D surface. This system has been used as a first approximation for the description of electron layers bound to the surface of liquid dielectrics and of inversion layers
in semiconductor physics~\cite{BausPR1980,FortovBook,KhrapakCPP2016}. 

There is another system of interacting particles, also referred to as the 2D OCP. Here, the interaction potential is defined via the 2D Poisson equation and scales {\it logarithmically} with distance.
The logarithmic potential, $\varphi(r)=-q^2\ln(r/L)$, where $L$ is an arbitrary scaling length, corresponds to the interaction of infinite charged filaments (and hence we use a different symbol to denote the charge). It is often used to model interactions between vortices in thin-film superconductors. The 2D OCP has received considerable attention~\cite{Caillol1982,deLeeuw1982,Choquard1983,Radloff1984},  because of various field-theoretical
models~\cite{BausPR1980}, and the existence of exact analytic solutions for some special cases~\cite{Jancovici1981,Alastuey1981}. 

The third system considered here is a 2D fluid with a dipole-like repulsive interaction. The dipolelike potential $\varphi(r)=\epsilon(\sigma/r)^3$ occurs in various 2D physical systems ranging from ions and colloidal particles trapped at various interfaces, colloidal particles in external electric fields, paramagnetic particles exposed to external magnetic fields, electrical charges placed in a flowing collisionless plasma, etc. Not surprisingly, these systems have also received considerable attention~\cite{vonGrunbergPRL2004,TeeffelenEPL2006,GoldenPRE2010,KhrapakPRE02_2018,KryuchkovJCP2019}.

The fourth system considered is the 2D fluid of repulsive Yukawa particles. The Yukawa potential (also known as the Coulomb screened or the Debye-H\"uckel potential) emerges when electrical interactions between charged particles immersed in a polarizable neutralizing medium are considered. This can be relevant in the context of colloidal and complex (dusty) plasma monolayers (also known as 2D plasma crystals and fluids~\cite{ThomasPRL1994,ChuPRL1994,ThomasNature1996,KonopkaPRL2000,HartmannPRE2005,RatynskaiaPRL2006,FortovUFN,FortovPR,YuPRE2024,YuPRE06_2024}). The Yukawa potential is $\varphi(r)=(Q^2/r)\exp(-r/\lambda)$, where $\lambda$ is the screening length. In the limit $\lambda\rightarrow \infty$ we recover the Coulomb OCP limit. In the opposite $\lambda\rightarrow 0$ limit, the interaction tends to that of hard disks in 2D, but this is not considered here.

The thermodynamics and dynamics of the systems considered are governed by the coupling parameter $\Gamma$ and the characteristic frequency $\omega_0$. For the Yukawa fluid, an additional screening parameter $\kappa=a/\lambda$ plays a role, where $a=(\pi n)^{-1/2}$ is the system-independent 2D Wigner-Seitz radius. The definitions are system-dependent and should be specified for convenience. For the 2D OCP with Coulomb interaction and Yukawa system, the coupling parameter is defined as the conventional Coulomb coupling parameter $\Gamma=Q^2/aT$. The 2D plasma frequency is $\omega_0=\sqrt{2\pi Q^2 n/ma}$~\cite{KhrapakJCP2018_1,KhrapakPoP03_2018}. For the 2D OCP with logarithmic potential, the coupling parameter is density independent, $\Gamma=q^2/T$~\cite{KhrapakCPP2016}, while the corresponding plasma frequency is $\omega_0=\sqrt{2\pi q^2 n/m}$~\cite{KhrapakPoP05_2016}. For the 2D fluid with dipole-like interaction, the coupling parameter is $\Gamma = (\epsilon/T)(\sigma/a)^3$, while the characteristic frequency is $\omega_0=\sqrt{2\pi n\epsilon\sigma^3/m a^3}$~\cite{KhrapakPRE02_2018}. 

Substituting these definitions into the equation (\ref{VibrationalEntropy}) allows us to present the excess entropy of 2D fluids in a {\it universal} simple generic form
\begin{equation}\label{sEx_gen}
s_{\rm ex}=1-\ln\Gamma-\ln\left(\frac{\omega_{\rm E}^2}{\omega_0^2}\right).
\end{equation}
Only the last term is system-specific and is responsible for some difference in the entropy of different fluids. 

\begin{table}
\caption{\label{Tab1} Fitting parameters $a$, $b$, and $c$ in Eq.~(\ref{sFit}) for 2D OCP with Coulomb ($1/r$) interaction, 2D OCP with logarithmic ($-\ln r$) interaction, and a 2D fluid with dipole-like ($1/r^3$) interaction. Based on the results reported in Refs.~\cite{KhrapakCPP2016,KhrapakPRE02_2018}}.
\begin{ruledtabular}
\begin{tabular}{crrrr}
System & $a$ & $b$ & $c$   \\ \hline
OCP ($1/r)$ & 0.231 & 2.798 & 0  \\
OCP ($-\ln r$) & 0.231 & 2.798 & 0.1469  \\
Dipole ($1/r^3$) & 0.2728 & 2.2357  & 0.381  \\
\end{tabular}
\end{ruledtabular}
\end{table}

Accurate simulation results for thermodynamic properties exist for all of the systems considered. Practical fitting formulas of varying complexity and accuracy have also been proposed, which allow an easy derivation of the excess Helmholtz free energy and the excess entropy. Here we use fitting formulas based on the molecular dynamics (MD)
results from Ref.~\cite{GannPRB1979} for the 2D OCP with the Coulomb interaction, the Monte Carlo (MC)~\cite{Caillol1982} and MD~\cite{deLeeuw1982} results for the 2D OCP with the logarithmic interaction, and MD results for a 2D system with dipole-like interactions~\cite{KhrapakPRE02_2018}. The fitting formulas for 2D OCP systems are summarized in Ref.~\cite{KhrapakCPP2016}, while the appropriate formula for the 2D fluid with dipole-like interactions can be found in Ref.~\cite{KhrapakPRE02_2018}. For the 2D Yukawa fluid, several relatively accurate expressions have been proposed for excess internal energy~\cite{KhrapakPoP08_2015,HuangPoP2017,KryuchkovJCP2017,CastelloPRE2021}. Since these serve as a basis for our present comparison, the details will be summarized below. 

For 2D Coulomb and logarithmic OCP as well as for dipole particles, numerical simulations have been used to fit the thermal component of the internal energy (additional to the static lattice sum also denoted as cold energy) by a universal 2D expression of the form $u_{\rm th}\simeq a\ln(1+b\Gamma)$~\cite{KhrapakCPP2016,KhrapakPRE02_2018,KryuchkovJCP2017}. This results in a quasi-universal expression for the excess entropy
\begin{equation}\label{sFit}
s_{\rm ex}\simeq a\ln(1+b\Gamma) +a{\rm Li}_2(-b\Gamma)+c,
\end{equation}
where ${\rm Li}_2(x)$ is a polylogarithm function and $a$, $b$, $c$ are fitting parameters. The values they take in fluids we consider, based on the results reported in Refs.~\cite{KhrapakCPP2016,KhrapakPRE02_2018} are summarized in Tab.~\ref{Tab1}. 

\begin{figure}
\includegraphics[width=7cm]{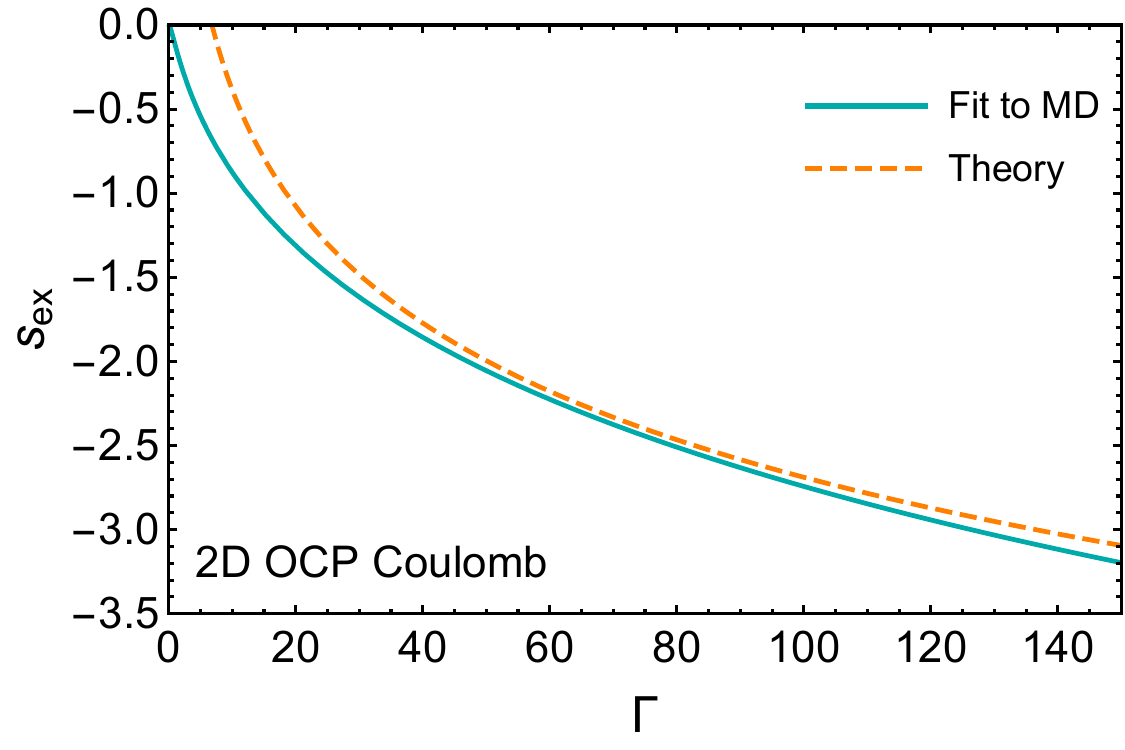}
\caption{(Color online) Excess entropy of the 2D OCP with Coulomb ($\propto 1/r$) interaction versus the coupling parameter $\Gamma$. The solid line corresponds to a fit based on MC simulation results~\cite{GannPRB1979}. The dashed line corresponds to the present theoretical model.  }
\label{Fig1}
\end{figure}

For the 2D Yukawa fluid, we use an equation of state proposed in Ref.~\cite{KryuchkovJCP2017}, which combines relative simplicity and accuracy (a detailed comparison between different equations of states for 2D Yukawa fluids can be found in Ref.~\cite{CastelloPRE2021}). The excess thermal energy is expressed as
\begin{equation}\label{fit_Yukawa}
u_{\rm th}(\kappa,\Gamma)=a(\kappa)\ln\left[1+b(\kappa)\Gamma^{s(\kappa)}\right]    
\end{equation}
The $\kappa$-dependent coefficients are $a(\kappa)=0.357 + 0.094\kappa$, $b(\kappa) = 1.655\exp(-0.769\kappa)$, and $s(\kappa) = 0.688 - 0.052\kappa$. The thermal contribution to the excess Helmholtz free energy can be computed from the thermodynamic relation 
\begin{equation}\label{integr_Helmholtz}
f_{\rm th}(\kappa,\Gamma)=\int_0^{\infty}\frac{u_{\rm th}(\kappa,\Gamma')}{\Gamma'}d\Gamma'.
\end{equation}
The problem here is that Eq.~(\ref{fit_Yukawa}) applies in the strongly coupled regime and is not designed to be accurate in the limit $\Gamma\rightarrow 0$. To mitigate this issue, we start the numerical integration from $\Gamma'=1$ in Eq.~(\ref{integr_Helmholtz}) and add the value $f_{\rm ex}(\kappa,1)$ evaluated from the second virial coefficient, just as was done in Ref.~\cite{KhrapakPPCF2015} for the 3D Yukawa fluid. The resulting expression is
\begin{equation}\label{integr_Helmholtz1}
f_{\rm th}(\kappa,\Gamma)=\int_1^{\infty}\frac{u_{\rm th}(\kappa,\Gamma')}{\Gamma'}d\Gamma' +f_{\rm th}(\kappa,1).
\end{equation}
Finally, since the configurational contribution to the excess energy is proportional to $\Gamma$~\cite{KryuchkovJCP2017}, it does not contribute to the excess entropy. The latter is obtained from
\begin{equation}\label{sYuk}
 s_{\rm ex}(\kappa,\Gamma)=u_{\rm th}(\kappa,\Gamma)-f_{\rm th}(\kappa,\Gamma).   
\end{equation}

\begin{figure}
\includegraphics[width=7cm]{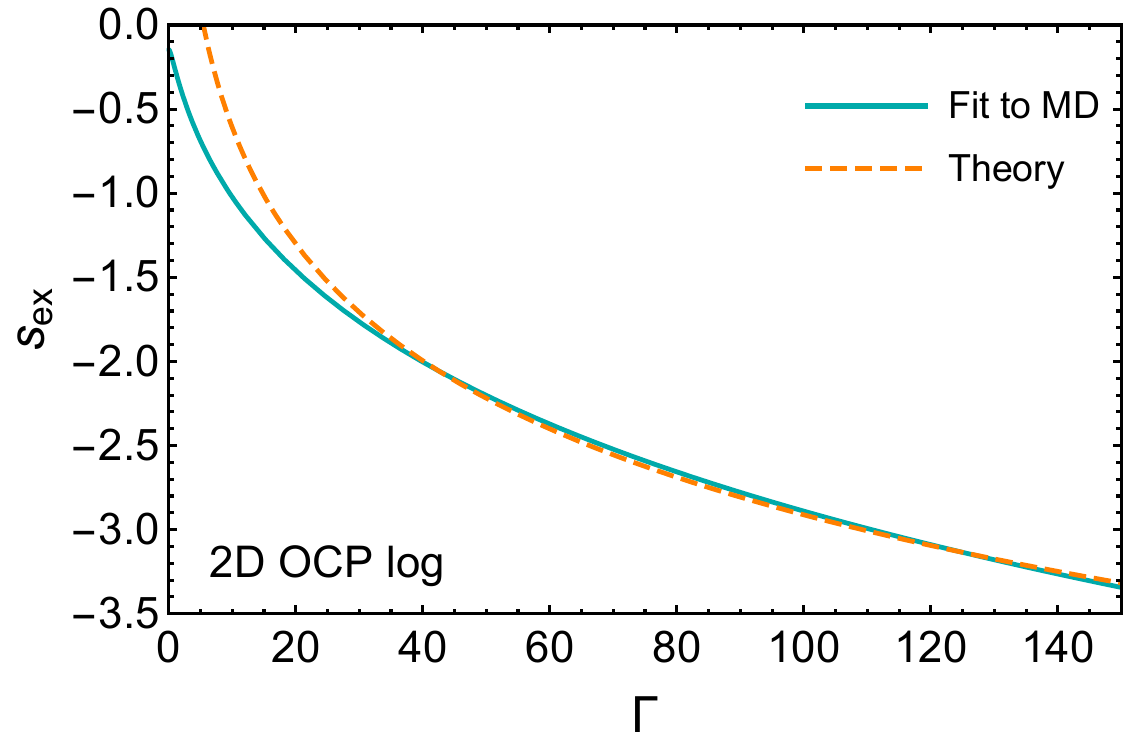}
\caption{(Color online) Excess entropy of the 2D OCP with logarithmic interaction ($\propto \ln(1/r)$) versus the coupling parameter $\Gamma$. The solid line corresponds to a fit based on MC and MD simulation results~\cite{Caillol1982,deLeeuw1982}. The dashed line corresponds to the present theoretical model. }
\label{Fig2}
\end{figure}

Now, a detailed comparison can be performed. The dependence of the excess entropy of the 2D OCP fluid with the Coulomb interaction potential on the coupling parameter $\Gamma$ is shown in Fig.~\ref{Fig1}. The solid curve shows the fit based on the numerical results, the dashed curve is plotted using Eq.~(\ref{sEx_gen}). In this case, the Einstein frequency can be estimated as a lattice sum for dipole-dipole ($\propto 1/r^3$) interactions $\omega_{\rm E}^2\simeq 0.4 \omega_0^2$~\cite{KhrapakPoP03_2018,DonkoJPCM2008}. This estimate is sufficiently accurate in the strong coupling regime, but loses accuracy as $\Gamma$ decreases. However, the vibrational paradigm itself is only applicable at strong coupling, so this does not represent any considerable problem. We observe in Fig~\ref{Fig1} that the theory compares well with the numerical results in the regime $\Gamma\gtrsim 50$.    In the vicinity of the fluid-solid phase transition at $\Gamma\simeq 140$ the theory slightly overestimates the numerical results.     

\begin{figure}
\includegraphics[width=7cm]{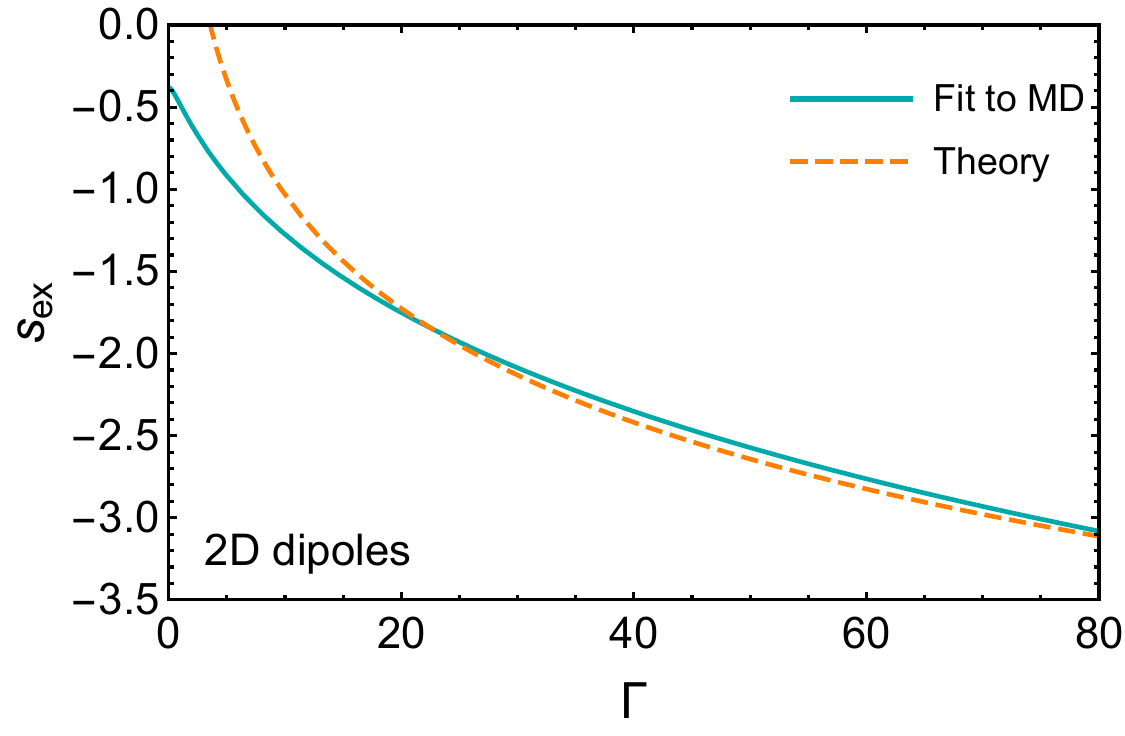}
\caption{(Color online) Excess entropy of the 2D fluid with dipole ($\propto 1/r^3$) interaction versus the coupling parameter $\Gamma$. The solid line corresponds to a fit based on MD simulation results~\cite{KhrapakPRE02_2018}. The dashed line corresponds to the present theoretical model.  }
\label{Fig3}
\end{figure} 

Similarly, Fig.~\ref{Fig2} shows a comparison between the vibrational approximation to excess entropy and the numerical data for 2D OCP with the logarithmic interaction potential. In this case the situation is simplified since the ratio of the Einstein to plasma frequency is fixed by the Kohn's sum rule~\cite{GoldenPoP2000,KhrapakPoP05_2016}, $\omega_{\rm E}^2=\omega_0^2/2$. The theory compares remarkably well with the simulations at $\Gamma\gtrsim 30$.         

\begin{figure*}
\includegraphics[width=18cm]{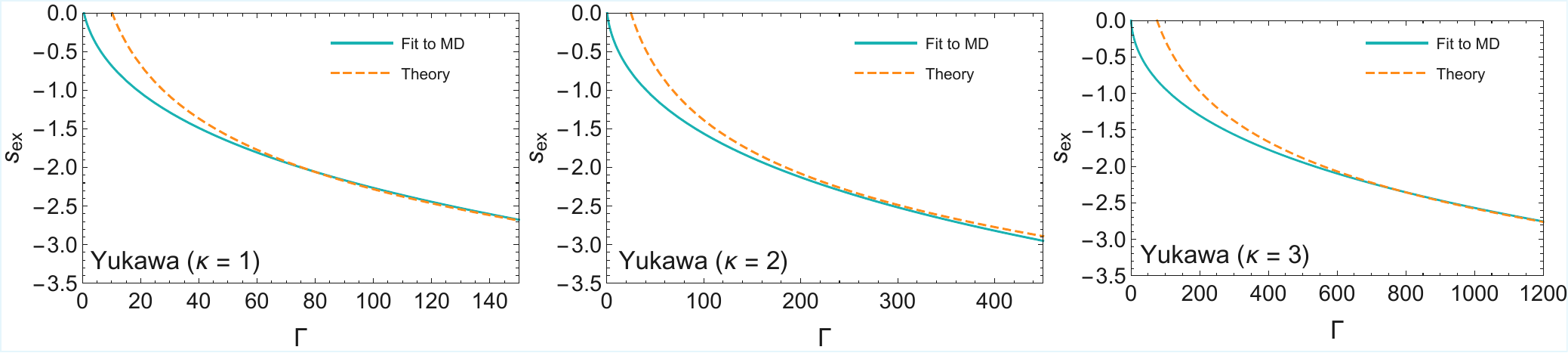}
\caption{(Color online) Excess entropy of the 2D Yukawa fluid versus the coupling parameter $\Gamma$ for three values of the screening parameter: $\kappa=1$ (left), $\kappa=2$ (center), and $\kappa=3$ (right). The solid lines correspond to a fit based on MD simulation results~\cite{KryuchkovJCP2017}. The dashed lines correspond to the present theoretical model.}
\label{Fig4}
\end{figure*}  
 
Figure~\ref{Fig3} presents a comparison between theory and numerical data for a 2D fluid with the dipole-like inverse cube interaction potential. The estimate of the Einstein frequency at strong coupling can be made using the results reported in Ref.~\cite{KhrapakPRE02_2018}. Specifically, $\omega_{\rm E}^2\simeq \tfrac{3}{2}\omega_0^2/R^3$, where $R\simeq 1.2523$ is the radius of the correlation hole expressed in units of the Wigner-Seitz radius (see Ref.~\cite{KhrapakPRE02_2018} for details). Numerically, we get $\ln(\omega_{\rm E}/\omega_0)^2\simeq -0.2695$.  The theory describes the numerical data well at $\Gamma\gtrsim 20$ up to the freezing point at $\Gamma\simeq 69$~\cite{KhrapakPRE02_2018}. 

Figure~\ref{Fig4} shows a comparison between the vibrational approximation to excess entropy and the numerical data for the 2D Yukawa fluid with $\kappa =1$, 2, and 3. For the 2D Yukawa fluid, the ratio $\omega_{\rm E}/\omega_0$ is only weakly dependent on $\Gamma$, but drops considerably as $\kappa$ increases (see Fig. 4 from Ref.~\cite{KhrapakPoP03_2018}). We use the values reported in Ref.~\cite{DonkoJPCM2008}, corresponding to the vicinity of the fluid-solid phase transition: $\omega_{\rm E}/\omega_0\simeq 0.5159$ ($\kappa=1$), $\simeq 0.3297$ ($\kappa=2$), and $\simeq 0.1893$ ($\kappa=3$). The agreement between the prediction of the vibrational model and the equation of state is again remarkably good at strong coupling. 

For Yukawa fluids we limit ourselves to the weak screening regime $\kappa\leq 3$ for the following reasons. The model of excess thermal energy of Eq.~(\ref{fit_Yukawa}) has only been verified and tested for $\kappa\lesssim 3$. It uses the fits for the Madelung (cold) energy and the melting coupling parameter that are only applicable for $\kappa\lesssim 3$~\cite{KryuchkovJCP2017}. The precision of other equations of state was tested in Ref.~\cite{CastelloPRE2021} only up to $\kappa=3.0$. For the Einstein frequency, we use the values reported in Ref.~\cite{DonkoJPCM2008} only for $\kappa = 1.0$, 2.0 and 3.0. A differential equation relating the Einstein frequency and the excess energy has been derived in Ref.~\cite{KhrapakPoP03_2018,KhrapakPoP12_2024_Erratum}, but the problem is again in the absence of an accurate equation of state at $\kappa>3.0$. Thus, unfortunately, there are no reliable results for both excess entropy and Einstein frequency to verify how adequate the current realization of the vibrational model behaves as $\kappa$ increases and the interaction potential steepens. 

\section{Discussion}

\subsection{Comparison between 2D and 3D fluids}

We have observed that for all model 2D fluids considered, the vibrational paradigm of atomic dynamics combined with the simple Einstein approximation with identical vibrational frequencies for all particles provides more than a satisfactory agreement with the numerical data on excess entropy. In reality, fluids can be characterized by some distribution of the oscillation frequencies, known as the vibrational density of states (VDOS). In a previous application of the vibrational picture to 3D fluids, a related term $\langle \ln (\omega^2/\omega_0^2)\rangle$ has been evaluated using the Debye-like model, assuming acoustic dispersion for the collective longitudinal and transverse modes~\cite{KhrapakJCP2021}. For very soft interactions, such as in the OCP fluid, this is not appropriate because the longitudinal mode is non-acoustic. In this case, the approximate dispersion relations can be used instead, without assuming their acoustic character~\cite{KhrapakJCP2021}. In particular, the use of a simple analytical approximation for the longitudinal and transverse collective modes of the strongly coupled 3D Yukawa fluid provides excellent precision in evaluating excess entropy within the vibrational paradigm~\cite{KhrapakPRE09_2024}. 
In view of the agreement already achieved, there is probably no need to further complicate the computation for 2D fluids beyond the simple Einstein approximation. At the same time, two important related questions deserve to be addressed. The first question is whether using dispersion relation (DR)-based averaging can improve or at least deliver the same level of accuracy for 2D fluids. This could be expected on the basis of our previous results for 3D fluids. The second question is whether the Einstein model works equally well for 3D fluids as it does for 2D fluids. If so, this would drastically simplify the entropy estimation in this case.   

To clarify these points, let us consider genuine OCP fluids in 2D (logarithmic potential) and 3D (Coulomb potential). These systems are chosen for three main reasons: (i) highly accurate thermodynamic data are available; (ii) the ratio $\omega_{\rm E}/\omega_0$ is fixed across the phase diagram ($\omega_{\rm E}/\omega_0=1/\sqrt{d}$, where $d=2,3$ is the dimensionality); (iii) simple yet accurate approximations for the dispersion relations of the longitudinal and transverse collective modes are available. 

We briefly recall that the 3D OCP fluid is described by the conventional Coulomb coupling parameter $\Gamma=Q^2/aT$, where $a=(4\pi n/3)^{-1/3}$ is the 3D Wigner-Seitz radius~\cite{BausPR1980,BrushJCP1966,IchimaruRMP1982}. The conventional plasma frequency is $\omega_0=\sqrt{4\pi Q^2 n/m}$. The excess entropy of 3D fluids within the vibrational model can be estimated from~\cite{KhrapakJCP2021}
\begin{equation}
s_{\rm ex}=\frac{3}{2}-\frac{3}{2}\left\langle \ln\frac{m\Delta^2\omega^2}{2\pi T} \right\rangle,   
\end{equation}
where $\Delta=n^{-1/3}$ in 3D. For the OCP fluid, this can be rewritten as
\begin{equation}
s_{\rm ex}=\frac{3}{2}\left(1-\ln\Gamma -\ln\left[2\left(\frac{4\pi}{3}\right)^{-1/3}\right]-\left\langle\ln\frac{\omega^2}{\omega_0^2}\right\rangle\right).
\end{equation}
The 2D analogue of Eq.~(\ref{sEx_gen}) reads
\begin{equation}\label{sEx2D}
s_{\rm ex}=1-\ln\Gamma-\left\langle\ln\frac{\omega^2}{\omega_0^2}\right\rangle.
\end{equation}
The term that needs to be evaluated is $\langle\ln(\omega^2/\omega_0^2)\rangle$. If the exact VDOS $g(\omega)$ were known, we would easily do that from
\begin{equation}
\left\langle\ln\frac{\omega^2}{\omega_0^2}\right\rangle = \int_0^{\infty} \ln\frac{\omega^2}{\omega_0^2}g(\omega) d\omega.  
\end{equation}
Unfortunately, in the fluid state, $g(\omega)$ is generally unknown, although serious efforts have been made to improve our understanding~\cite{RabaniJCP1997, ZacconePNAS2021,BaggioliPRE2021,StamperJPCL2022,SchirmacherPRE2022,BaggioliPRE2023}. The simplest approximation adopted here for 2D fluids is the Einstein approximation $g(\omega)\propto \delta(\omega-\omega_{\rm E})$. It obviously leads to $\langle\ln(\omega/\omega_0)^2\rangle=\ln(\omega_{\rm E}^2/\omega_0^2)$, independently of spatial dimensionality.

An alternative approximation is to express the average as a sum over normal modes with different wave vectors and polarizations and to replace these sums by integration in the $k$-space, using $\frac{1}{N}\Sigma_{\bf{k}}\rightarrow \frac{1}{n}\int d{\bf{k}}/(2\pi)^d$, where $d$ again denotes spatial dimensionality~\cite{LandauStatPhys,KhrapakPRE02_2018,KhrapakJCP2021,KhrapakPRR2020}. This leads to the 3D expression~\cite{KhrapakJCP2021}
\begin{equation}\label{Av3}
 \left\langle\ln\frac{\omega^2}{\omega_0^2}\right\rangle=\frac{2}{9\pi}\int_0^{q_{\rm max}}q^2dq\left[\ln\frac{\omega_l(q)^2}{\omega_0^2}+2\ln\frac{\omega_t(q)^2}{\omega_0^2}\right],   
\end{equation}
which accounts for one longitudinal and two transverse modes. Here, $q_{\rm max} = (9\pi/2)^{1/3}\simeq 2.418$ is the cutoff wave number, which ensures that a quantity that does not depend on $q$ remains invariant upon averaging. In 2D geometry, we get~\cite{KhrapakPRE02_2018} 
\begin{equation}\label{Av2}
 \left\langle\ln\frac{\omega^2}{\omega_0^2}\right\rangle=\frac{1}{4}\int_0^{q_{\rm max}}qdq\left[\ln\frac{\omega_l(q)^2}{\omega_0^2}+\ln\frac{\omega_t(q)^2}{\omega_0^2}\right],   
\end{equation}
accounting for one longitudinal and one transverse mode. In the 2D case, we get $q_{\rm max} = 2$.

\begin{figure}
\includegraphics[width=7cm]{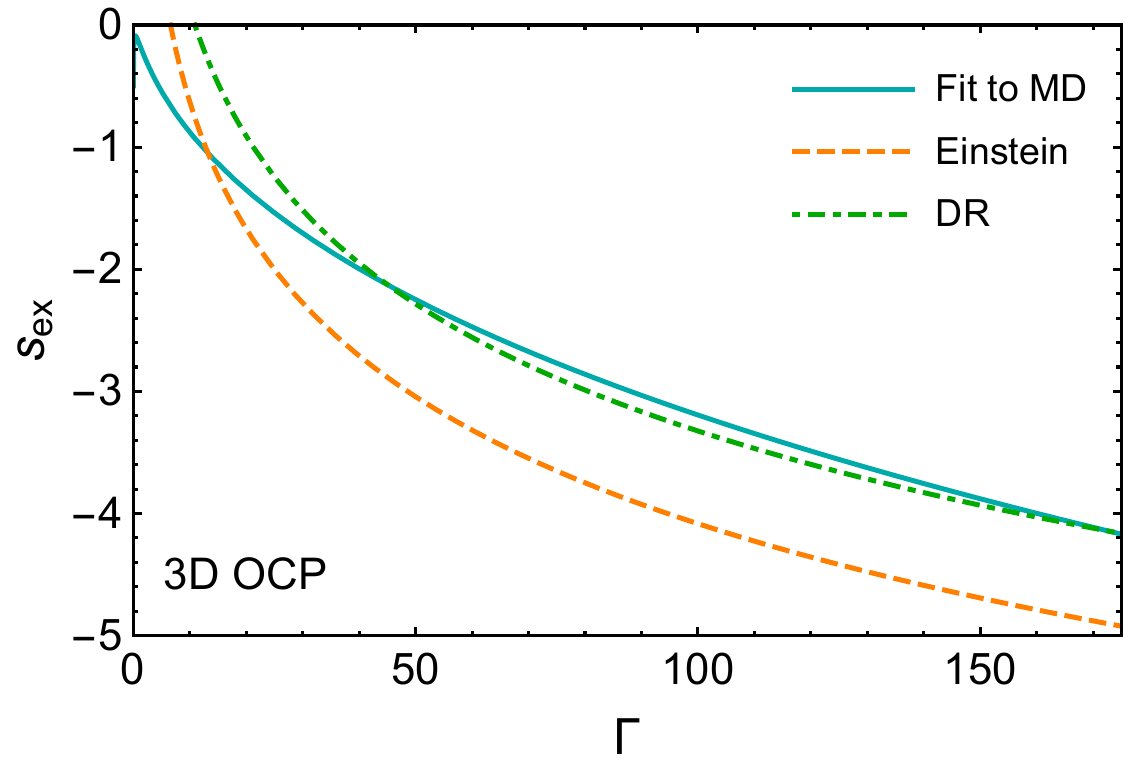}
\caption{(Color online) Excess entropy of the 3D OCP fluid versus the coupling parameter $\Gamma$. The solid line corresponds to a fit based on MD simulation results~\cite{KhrapakCPP2016}. The dashed line corresponds to the vibrational model with Einstein approximation. The dash-dotted curve is obtained using DR-based averaging as explained in the text.}
\label{Fig5}
\end{figure} 

\begin{figure}
\includegraphics[width=7cm]{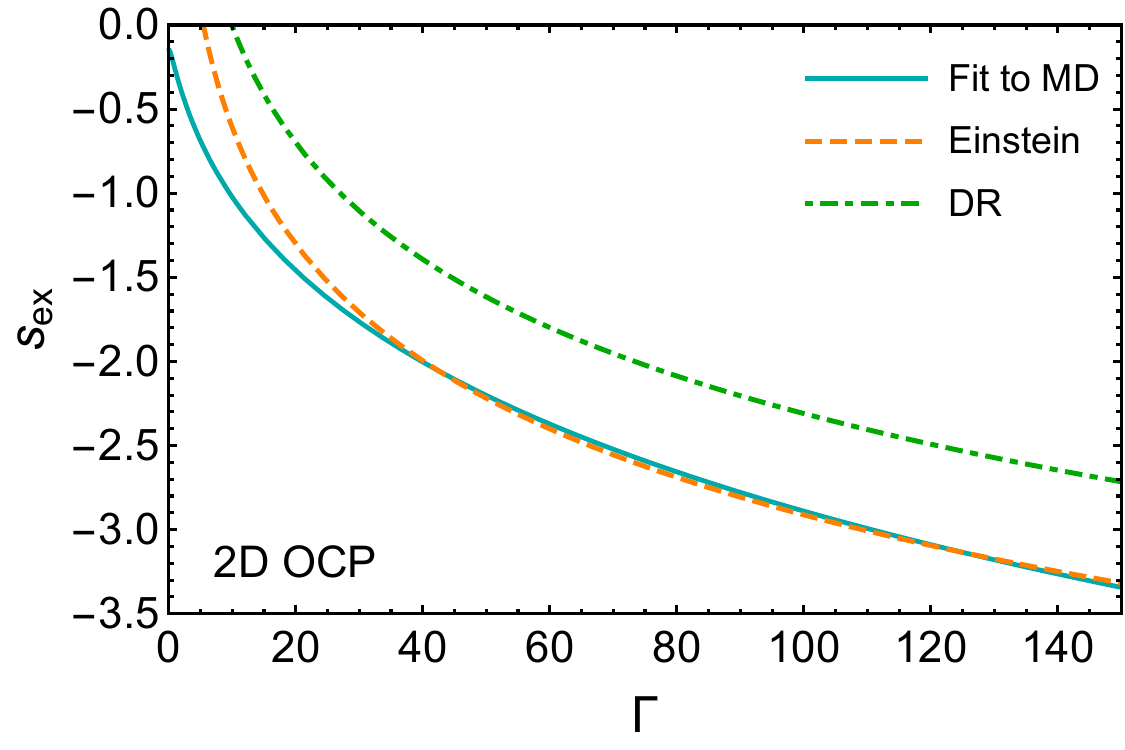}
\caption{(Color online) Excess entropy of the 2D OCP fluid versus the coupling parameter $\Gamma$. The solid line corresponds to a fit based on MC and MD simulation results~\cite{Caillol1982,deLeeuw1982}. The dashed line corresponds to the vibrational model with Einstein approximation. The dash-dotted curve is obtained using DR-based averaging as explained in the text.}
\label{Fig6}
\end{figure} 

Simple yet accurate dispersion relations for $\omega_l(q)$ and $\omega_t(q)$ appropriate for 3D and 2D OCP fluids are provided in the Appendix. These have been used to perform the numerical integration in Eqs.~(\ref{Av3}) and (\ref{Av2}). The results are shown in Figs.~\ref{Fig5} and \ref{Fig6}. Figure~\ref{Fig5} shows the calculation for the 3D OCP. It demonstrates that DR-based averaging provides good accuracy at $\Gamma\gtrsim 50$ as already reported in Ref.~\cite{KhrapakJCP2021}. Note that DR-based averaging is generally quite appropriate in 3D and provides reliable results not only for excess entropy~\cite{KhrapakJCP2021,KhrapakPRE09_2024}, but also for the thermal conductivity coefficient~\cite{KhrapakPRE01_2021,KhrapakPoP08_2021,KhrapakPRE12_2023}, and for the Stokes-Einstein relation between the self-diffusion and viscosity coefficients~\cite{ZwanzigJCP1983,KhrapakPRE10_2021,KhrapakMolecules12_2021}. Einstein approximation does not work so well in 3D and results in $\simeq 20\%$ underestimation of the reduced excess entropy. On the contrary, for 2D OCP the Einstein approximation is very accurate, as we already know -- see Figs.~\ref{Fig1} --\ref{Fig4}. As demonstrated in Fig.~\ref{Fig6}, the DR-based approximation overestimates the excess reduced entropy by some $\simeq 20\%$. We therefore have to conclude that although the Einstein approximation is generally quite accurate for 2D fluids, for 3D fluids it does not perform so well, and the DR-based averaging is more reliable and preferable in this case.     

It may seem puzzling that using more detailed information about collective modes, contained in dispersion relations, makes the agreement worse in 2D geometry compared to perhaps the simplest possible Einstein model. Although this apparently deserves further detailed consideration, we can now point out two potential issues with DR-based implementation for 2D systems. First and perhaps most important is that this approach neglects the so-called $q$-gap -- the zero-frequency portion of the dispersion relation at low $q$ also known as the propagation gap in the
dispersion relation of the transverse collective mode~\cite{HansenBook,OhtaPRL2000,MurilloPRL2000,NosenkoPRL2006,GoreePRE2012,BrykPRE2014,BolmatovPCL2015,TrachenkoRPP2015,YangPRL2017,KhrapakJCP2019,KryuchkovSciRep2019,KryuchkovJCP2021,KhrapakPRE11_2024,BrykPRE2025}. Although the $q$ gap exists in both 3D and 2D fluids, its relative importance should be higher in 2D, due to the geometrical factor $q^2$ in Eq.~(\ref{Av3}) as compared to the geometrical factor $q$ in Eq.~(\ref{Av2}). This makes the integral in 3D much less sensitive to the dispersion relations at small $q$. Perhaps the most straightforward way to verify this is to plot the functions under the integrals in Eqs.~(\ref{Av3}) and (\ref{Av2}), which clearly reveal differing sensitivities of the integrals to the accuracy of the dispersion relations at long wavelengths (small $q$). Another effect not taken into account in our derivation is the hybridization and anti-crossing of the longitudinal and transverse modes, which has recently been investigated theoretically and experimentally~\cite{KryuchkovJPCL2019,YakovlevJPCL2020}. However, this effect is probably not so important because it is most pronounced in the vicinity of the roton minima, beyond the cutoff.

\subsection{Applicability}

Returning to the results obtained for 2D fluids, the applicability domain should be discussed. First, it should be noted that the potentials considered in this paper are all repulsive, relatively soft and long-ranged, including the Yukawa potential, as only the regime $\kappa\leq 3.0$ is involved. This probably explains why the model works so well in all these cases. The important assumption of the model -- the harmonic potential of Eq.~(\ref{harmonic}) is appropriate for sufficiently soft interactions. At the same time, it might become less appropriate for steep interaction potentials, and the model could fail at some point. In the end, the Einstein frequency diverges in the extreme anharmonic limit of hard disks. Yukawa fluids with large screening parameters ($\kappa>3$) could help locate such a failure in future work. In addition, for 3D fluids, the presence of attractive branch is known to reduce the accuracy of the vibrational model~\cite{KhrapakJCP2021}, and it would be interesting to check whether the same occurs in 2D.

The onset of applicability of the vibrational approximation in terms of the coupling parameter $\Gamma$ varies from system to system, especially considering the 2D Yukawa fluid. However, in terms of excess entropy itself, the applicability condition can be uniquely formulated as $s_{\rm ex}\lesssim -2.0$. This coincides with the onset of vibrational dominance of the atomic dynamic of dense fluids, also known as the transition to the ``rigid-fluid regime '' in 3D~\cite{KhrapakPhysRep2024,KhrapakPRE07_2025}. This is also the applicability condition of the vibrational model of entropy applied to 3D fluids in Ref.~\cite{KhrapakJCP2021}. Thus, the applicability condition formulated in terms of entropy does not seem to depend on the dimensionality of the system.

\subsection{Excess entropy at freezing}

Another important reference point is the fluid-solid phase transition and the freezing point, defined as the fluid boundary of the fluid-solid coexistence region (which can be rather narrow for soft repulsive potentials and vanishes for OCP). From the theoretical model developed here we obtain $s_{\rm ex}\simeq -3.1$ for the OCP with Coulomb interaction potential (assuming $\Gamma_{\rm fr}\simeq 135$~\cite{KhrapakCPP2016}), $s_{\rm ex}\simeq -3.2$ for the OCP with the logarithmic interaction potential (assuming $\Gamma_{\rm fr}\simeq 135$~\cite{KhrapakCPP2016}), and $s_{\rm ex}\simeq -2.9$ for the 2D dipole system (assuming $\Gamma_{\rm fr}\simeq 69.2$). For the 2D Yukawa fluid, the excess entropy at freezing slowly increases with $\kappa$:  $s_{\rm ex}\simeq -2.86$ ($\kappa=1$), $-2.81$ ($\kappa=2$) and $-2.71$ ($\kappa=3$), as obtained from the equation of state in Ref.~\cite{KryuchkovJCP2017} and freezing line from Ref.~\cite{YuPRE06_2024}. In 3D fluids, freezing is known to occur at a lower excess entropy ($s_{\rm ex}\simeq -4$)~\cite{KhrapakJCP2021,RosenfeldPRE2000,KhrapakJCP2022_1,KhrapakJPCL2022,HeyesJCP2025} and there is a good explanation for this based on the Lindemann melting criterion~\cite{KhrapakPRE09_2024}. Note that the conventional Lindemann melting criterion is not applicable to 2D systems, but its modifications have been proposed~\cite{KhrapakPRR2020} and relations to the celebrated Berezinskii-Kosterlitz-Thouless-
Halperin-Nelson-Young (BKTHNY) theory~\cite{Kosterlitz2017} have also been discussed~\cite{KhrapakJCP2018}. 

\subsection{Heat capacity}

It should be noted that, despite the relatively good accuracy demonstrated by Eq.~(\ref{sEx_gen}), it does not provide access to fine details of the specific heat behavior. The reduced excess component of the isochoric heat capacity can be expressed as $c_{\rm v}^{\rm ex}=-\Gamma(\partial s_{\rm ex}/\partial \Gamma)$~\cite{KhrapakPRE09_2024}. From Eq.~(\ref{sEx_gen}) it follows that $c_{\rm v}^{\rm ex}= 1$ as long as the dependence of $\omega_{\rm E}/\omega_0$ on $\Gamma$ is neglected. This is the expected harmonic 2D results, analogue of the conventional Dulong-Petit law. In reality, $c_{\rm v}^{\rm ex}$ should gradually decrease from $\simeq 1$ near the fluid-solid phase transition to zero in the ideal gas limit (although our model is not applicable in this regime). Some initial decrease in $c_{\rm v}^{\rm ex}$ could possibly be explained by introducing the dependence of $\omega_{\rm E}/\omega_0$ on $\Gamma$. However, this explanation is not valid in general: For the 2D OCP with logarithmic potential, the identity $\omega_{\rm E}\equiv\omega_0/\sqrt{2}$ is valid. Thus, the current implementation of the vibrational model is unable to resolve the fine details of $c_{\rm v}$.  On the contrary, Eqs.~(\ref{sFit}) and (\ref{sYuk}) can be used for this purpose.   

\subsection{Pair entropy}

It is worth recalling that the pair contribution to entropy is a good approximation for the total entropy in 3D fluids near freezing~\cite{Baranyai1989,LairdPRA1992}. In particular, the residual multiparticle entropy (RMPE), $s_{\rm ex} -s_2$, is found to vanish in the vicinity of the fluid-solid phase transition in 3D~\cite{GiaquintaPhysA1992,GiaquintaPRA1992,SaijaJCP2006}. Recent results~\cite{KlumovResPhys2020} have demonstrated that $s_2$ and $s_{\rm ex}$ are not very close in 2D fluids near solidification, where the pair entropy overestimates the total entropy (i.e. $s_2$ appears more negative). In fact, RMPE vanishes somewhat prior to the freezing transition in 2D~\cite{KlumovResPhys2020,SaijaJCP2000}, so that $s_2$ and $s_{\rm ex}$ can deviate considerably at freezing. 

\subsection{Potential applications}

An important potential application of these results is related to the kinetic characterization of strongly coupled systems. To be concrete, let us consider 2D complex (dusty) plasmas -- systems of charged macroparticles immersed in the neutralizing plasma medium~\cite{TsytovichUFN1997,FortovUFN,FortovPR,FortovBook,ShuklaRMP_2009,MorfillRMP_2009,BonitzRPP2010}. Two-dimensional strongly coupled plasma crystals and fluids are commonly formed in the sheath area above the lower electrode of a radio-frequency discharge~\cite{ThomasPRL1994,ChuPRL1994,ThomasNature1996,NosenkoPRL2012}. At strong coupling, the vibrational dominance regime is naturally realized. The trajectory of macroparticles can be routinely measured with high spatial and temporal resolution using video microscopy. The interparticle separation, temperature, and Einstein frequency can be determined with a relatively high accuracy~\cite{WongIEEE2018}. This gives direct access to the fluid excess entropy by means of Eq.~(\ref{VibrationalEntropy}). The solid entropy can be estimated in a similar way by just eliminating the communal entropy term.  

Knowledge of the excess entropy opens up new perspectives in the study of strongly coupled 2D complex plasma systems. It is common to characterize complex plasma systems in terms of dimensionless coupling and screening parameters $\Gamma$ and $\kappa$, assuming that the Yukawa potential is a good approximation for interparticle interactions. Previously, experimental maps of $\Gamma$ and $\kappa$  measured at individual particle resolution have been reported~\cite{KnapekPRL2007}.  Experimental maps of excess entropy at the individual particle level can be even more helpful, because by virtue of isomorph theory, phase diagrams of Roskilde-simple (R-simple) systems are essentially one-dimensional in terms of $s_{\rm ex}$. For R-simple systems there exist so-called isomorphs in the $(n,T)$ phase diagram along which certain properly reduced structural, dynamical, and thermodynamic properties are invariant to a good approximation~\cite{GnanJCP2009,DyreJPCB2014,DyreJPCM2016}. Excess entropy is one of the properties that is approximately invariant along isomorphs and is conventionally used as a system control parameter. This is particularly relevant in view of the Rosenfeld's excess entropy scaling of transport coefficients~\cite{RosenfeldPRA1977,DyreJCP2018}. Strongly coupled Yukawa systems do belong to the R-simple class~\cite{VeldhorstPoP2015,YuPRE06_2024}, and this makes excess entropy a particularly insightful quantity. 

A straightforward application of the model would be to produce a {\it local} map of the Lindemann melting criterion. For 2D solids, the Lindemann criterion can be formulated using statistical mechanics arguments and expressed with the help of excess entropy~\cite{KhrapakPRR2020}. As suggested in Ref.~\cite{KnapekPRL2007}, this would allow us to identify the characteristic patterns of the caged particle motion in the vicinity of the melting transition and the role played by dynamical heterogeneity (dislocations, defects, grain boundaries). This seems particularly relevant since the 2D melting scenario can strongly depend on the softness of the interaction between particles and therefore be system dependent~\cite{KapferPRL2015}. Additionally, local maps of excess entropy in strongly coupled 2D fluids would provide an excellent opportunity to investigate scaling of dynamical phenomena with excess entropy. 

It is often considered that the Yukawa potential represents a reasonable approximation for particle-particle interactions in 2D complex plasmas~\cite{KonopkaPRL2000,KompaneetsPoP2007}. However, modifications have also been predicted, in particular due to the effect of directional ion motion in the sheath region of an rf discharge~\cite{VladimirovPRE1995,VladimirovPoP1996,KompaneetsPoP2007,KompaneetsPRL2016,KompaneetsPRE2016}. It is important to mention that the excess entropy estimated using Eq.~(\ref{VibrationalEntropy}) is {\it not based} on any assumption about the interaction potential. The Einstein frequency, interparticle separation, and particle kinetic temperature can all be measured experimentally, and thus ``true'' excess entropy is available. In contrast, if there are good grounds to believe that the actual interaction potential between the particles is close to Yukawa-like shape, then a comparison between experimentally measured $s_{\rm ex}$ and theoretical Eq.~(\ref{sYuk}) gives access to dimensionless parameters $\Gamma$ and $\kappa$. From this information, important complex plasma parameters, such as particle charge and plasma screening length, can be estimated. This can represent another useful {\it in situ} diagnostic tool for complex plasma parameters. 

In addition to complex plasmas, there are other systems that can potentially benefit from this model. The use of tunable interactions in 2D colloidal suspensions is a very active area of research. In particular, the use of rapidly rotating electric and magnetic fields is among the most promising due to their technological flexibility and the ability to change them {\it in situ}~\cite{KryuchkovJCP2019}. Various implementations of tunable interactions have been explored from both theoretical and experimental perspectives; see, e.g. Refs.~\cite{CoughlanJCP2017, DuSoftMat2017, YakovlevSciRep2017, KomarovSoftMat2018, KomarovSoftMat2020}. Estimating Einstein frequency experimentally may not be so easy in colloidal suspensions because of much stronger damping compared to complex plasma. However, if the pair interaction potential is known, the Einstein frequency can be calculated from structural information. This would again enable a straightforward connection to entropy and related thermodynamic properties without requiring thermodynamic integration.

\section{Conclusion}

To conclude, a simple vibrational approach to the entropy of dense 2D fluids based on the Einstein model is presented. The approach demonstrates good agreement with the available numerical results for four important systems: 2D OCP with the Coulomb interaction potential, 2D OCP with the logarithmic interaction potential, 2D fluid with isotropic dipole-like ($1/r^3$) repulsion, and 2D Yukawa fluid. Several important related aspects are briefly discussed, including the limits of applicability, relevance to 3D fluids, and connections to other freezing indicators in 2D systems. Potential practical applications of the model are briefly discussed.



\appendix

\section{Dispersion relations}

\begin{figure}
\includegraphics[width=7cm]{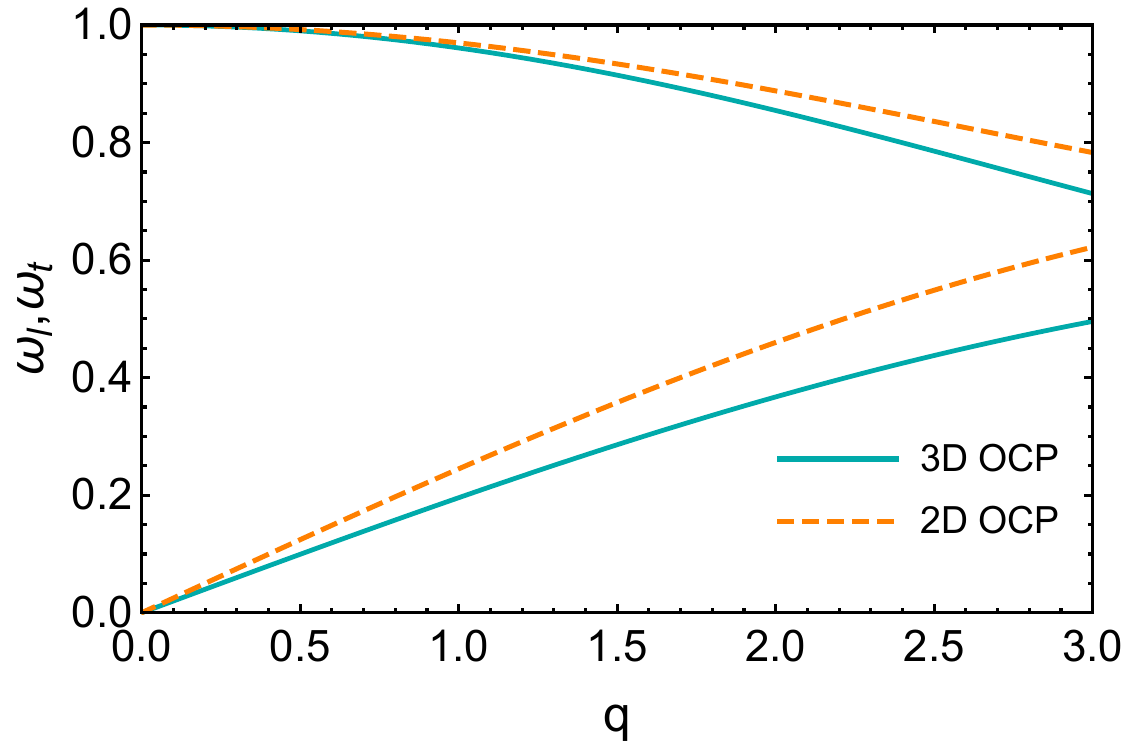}
\caption{(Color online) Dispersion relations of the longitudinal ($\omega_l$) and transverse ($\omega_t$) modes in the 3D strongly coupled OCP fluid (solid curves) and 2D strongly coupled OCP fluid (dashed curves). }
\label{Fig7}
\end{figure}

Simple yet accurate expressions for the longitudinal and transverse dispersion relations were derived in Refs.~\cite{KhrapakPoP2016,KhrapakPoP05_2016} using the quasi-localized charge approximation (QLCA) combined with a simple step-wise model of the radial distribution function $g(r)$. Their precision was compared with the results from numerical simulations in these papers as well as in Refs.~\cite{KhrapakAIPAdv2017,KhrapakIEEE2018}. The agreement with various numerical data is generally good at strong coupling and in the long-wavelength regime. For completeness, we provide these expressions in the following. For the 3D OCP fluid, the dispersion relations are~\cite{KhrapakPoP2016}
\begin{equation}\label{L2}
\frac{\omega_l^2(q)}{\omega_0^2}=\frac{1}{3}-\frac{2\cos Rq}{R^2q^2}+\frac{2\sin Rq}{R^3q^3} 
\end{equation}
and
\begin{equation}\label{T2}
\frac{\omega_t^2(q)}{\omega_0^2}=\frac{1}{3}+\frac{\cos Rq}{R^2q^2}-\frac{\sin Rq}{R^3q^3}. 
\end{equation}   
Here, $R\simeq 1.09545$ is the reduced radius of the correlation hole. The Kohn sum rule is automatically satisfied, $\omega_l^2+2\omega_t^2=\omega_{0}^2$. Since $\omega_l=\omega_t=\omega_{\rm E}$ in the individual particle limit $q\rightarrow \infty$ we also have $\omega_{\rm E}=\omega_0/\sqrt{3}$.
For the 2D OCP fluid the dispersion relations are~\cite{KhrapakPoP05_2016} 
\begin{equation}\label{SQLCA1}
\frac{\omega_l^2(q)}{\omega_0^2}= \frac{1}{2}+\frac{J_1(q)}{q},
\end{equation}
\begin{equation}\label{SQLCA2}
\frac{\omega^2_t(q)}{\omega_0^2}= \frac{1}{2}-\frac{J_1(q)}{q},
\end{equation}
where $J_1(x)$ is the Bessel function of the first kind. The 2D version of Kohn's sum rule $\omega_l^2+\omega_t^2=\omega_{0}^2$ is again automatically satisfied; the Einstein frequency is fixed $\omega_{\rm E}=\omega_0/\sqrt{2}$. The dispersion relations are plotted in Fig.~\ref{Fig7}. Note that in this approximation the $q$-gap, which is the zero-frequency portion of the transverse mode dispersion relation at low $q$ -- a general property of the liquid state -- is not reproduced.

\bibliography{SE_Ref}

\end{document}